\documentclass{article}
\usepackage{spconf,amsmath,graphicx}
\usepackage{mathrsfs,multicol,multirow,cite}
\usepackage{color}
\usepackage{url}
\usepackage{hyperref}
\usepackage{balance}
\usepackage{cite}

\title{Enhancing the Intelligibility of Cleft Lip and Palate Speech using\\ Cycle-consistent Adversarial Networks}

\name{Protima Nomo Sudro$^1$, Rohan Kumar Das$^3$, Rohit Sinha$^1$, S R Mahadeva Prasanna$^{2}$}
\address{
  $^1$Indian Institute of Technology Guwahati, Guwahati, India\\
  $^2$Indian Institute of Technology Dharwad, Dharwad, India\\
  $^3$National University of Singapore, Singapore}
\begin{document}

\maketitle
\begin{abstract}
Cleft lip and palate (CLP) refer to a congenital craniofacial condition that causes various speech-related disorders. As a result of structural and functional deformities, the affected subjects' speech intelligibility is significantly degraded, limiting the accessibility and usability of speech-controlled devices. Towards addressing this problem, it is desirable to improve the CLP speech intelligibility. Moreover, it would be useful during speech therapy. In this study, the cycle-consistent adversarial network (CycleGAN) method is exploited for improving CLP speech intelligibility. The model is trained on native Kannada-speaking childrens' speech data. The effectiveness of the proposed approach is also measured using automatic speech recognition performance. Further, subjective evaluation is performed, and those results also confirm the intelligibility improvement in the enhanced speech over the original.
   
\end{abstract}
\begin{keywords}
	CLP speech, intelligibility, CycleGAN, enhancement, speech disorder
\end{keywords}

\section{Introduction}

The individuals with cleft lip and palate (CLP) suffer from speech disorders due to velopharyngeal dysfunction, oro-nasal fistula, and mislearning~\cite{kummer2013cleft}. As a result, children with CLP may show different speech deviations such as hypernasality, articulation errors, nasal air emission, and voice disorders, and all these factors have an impact on the speech intelligibility~\cite{henningsson2008universal}. In the context of pathological speech, speech intelligibility is closely related to the notion of speech understandability, where it is defined as the degree to which the listener could understand the spoken message~\cite{whitehill2002assessing,henningsson2008universal}.

Hypernasality is a resonance disorder, and the presence of such nasal resonances during speech production has an excessively perceptible nasal quality~\cite{grunwell2001speech}. In addition, the vowels are nasalized, and the nasal consonants tend to replace the obstruents (i.e., stops, fricatives, and affricates)  due to severe hypernasality, all of which affect speech intelligibility~\cite{kummer2013cleft,scipioni2009intelligibility,maier2006intelligibility}. Besides hypernasality, the CLP speech intelligibility is also affected by deviant articulation patterns such as weak and nasalized consonants, glottal stops, pharyngeal and velar substitutions~\cite{peterson2001cleft,kummer2013cleft,moore1975phonetic}. The nasal air emission consists of an additional noise source, which becomes a part of the generated speech signal and influences the perceptivity of listeners~\cite{kummer2013cleft,kalita2019nasal}. The voice disorders include hoarseness and soft voice syndrome~\cite{peterson2001cleft}, but they may or may not affect the CLP speech intelligibility.

The advancements in technology have lead to various speech-based applications such as automatic speech recognition (ASR) and language identification to ease our daily lifestyles. However, people suffering from pathological speech cannot use such technologies effectively as the models of those systems are trained using normal speech~\cite{schuster2006evaluation,vucovich2017automated}. A study in~\cite{pradhan2018accessibility} reported an analysis of the use of speech-controlled devices for people with speech-based disorders. A few studies also investigated the ability of the speech assistants such as Siri, Google Assistant, Alexa, and Cortana to recognize speech from individuals with amyotrophic lateral sclerosis (ALS) induced dysarthria~\cite{ballati2018assessing,ballati2018hey}. 

The above studies show that people with different speech disabilities face many challenges using the latest speech-based technologies. However, many prefer to use speech-enabled services as one can perform a multitude of everyday tasks with less effort. One way to assist such people in using speech-enabled devices can be done by retraining the existing models, including different pathological speech types. However, the lack of a large amount of data for such cases compared to normal speech may be an obstacle. Further, retraining such models with pathological speech may affect the performance of those systems with normal speech. This shows the importance of improving the intelligibility of disordered speech to serve such needs. We consider the case of enhancement of the intelligibility of CLP speech.


The improvement of CLP speech intelligibility can be achieved clinically through surgery, prosthesis, and therapy. However, the surgical intervention may not result in functional correction of CLP speech, and deviant speech may persist even after surgery. In general, the speech-language pathologists (SLPs) assist the patients by showing the discrimination between the disordered and correct speech~\cite{kummer2013cleft}. Further, an SLP creates an awareness of the disorder by simulating the misarticulated speech sound and presenting it to the individual along with correct speech sounds~\cite{peterson2001cleft}. During this kind of speech therapy, the individuals with CLP learn about the perceptual contrast between correct and distorted sounds, which they try to rectify accordingly. Along a similar direction, we believe automatic systems can be built to correct CLP speech intelligibility towards that of normal speech, which motivates the current work.

\subsection{Related Works}

In the literature, various approaches are proposed for improving the intelligibility of different kinds of pathological speech. One of these corresponds to dysarthric speech modification based on acoustic transformation and spectral modification using the Gaussian mixture model (GMM)~\cite{rudzicz2013adjusting,kain2007improving}. The studies~\cite{bi1997application,liu2006enhancement} for alaryngeal speech enhancement include the transformation of speech by enhancing formants using chirp Z-transform and perceptual weighting techniques to adapt the subtraction parameters that effectively reduce the radiated noise of electrolaryngeal speech. Some other studies in~\cite{nakamura2012speaking,doi2013alaryngeal} improved the quality of electrolaryngeal speech using a speaking-aid system based on voice conversion (VC) method and one-to-many eigenvoice conversion. Similarly, the statistical approaches are exploited in~\cite{toda2012statistical} to enhance the body-conducted unvoiced speech for silence communication.

In~\cite{kong2012development}, the frequency lowering system and phoneme-specific enhancement were proposed for enhancing the intelligibility of degraded speech. A few studies also reported speech intelligibility enhancement for individuals with articulation disorders, glossectomy patients' speech using VC technique~\cite{fu2017joint,murakami2018naturalness}. The studies in~\cite{lai2019multi,biadsy2019parrotron} performed speech enhancement to reduce background noise in hearing aid devices for improving the intelligibility and naturalness for deaf speakers by adapting a pre-trained normalization model.

A VC method transforms the speech signal of one speaker into another while preserving the linguistic information~\cite{stylianou1998continuous,vcc2020summary}. Besides its application in pathological speech enhancement~\cite{kain2007improving,tanaka2016enhancing,fu2017joint,chen2018generative,yang2019self}, it has various other potential applications such as customizing audiobook and avatar voices, dubbing, computer-assisted pronunciation training, and voice restoration after surgery~\cite{turk2006robust,oyamada2017non}. In this regard, we plan to explore VC methods to transform the distorted CLP speech into more intelligible speech. 

The prior works show the use of GMM based VC and non-negative matrix factorization (NMF) based VC for the improvement of various pathological speech~\cite{kain2007improving,fu2017joint,sudro2019modification}. In this kind of method, temporally aligned parallel source (pathological speech) and target (normal/non-pathological speech) are required for training. However, collecting a large amount of pathological data such as CLP speech and creating a parallel corpus is challenging. This projects non-parallel VC methods more suitable for the current study. The cycle-consistent adversarial network (CycleGAN) is one of the state-of-the-art non-parallel VC methods that has shown its effectiveness for various applications~\cite{kaneko2018cyclegan,yeh2018rhythm,fang2018high}. Therefore, we consider CycleGAN to improve the intelligibility of the CLP speech in this work. We also study an NMF based method for a comparative study and perform objective and subjective analysis for speech intelligibility. This works' contribution lies in improving CLP speech intelligibility with CycleGAN to help people with pathological speech use speech-enabled devices.

The remaining paper is organized as follows. The CycleGAN system for enhancing speech intelligibility is discussed in Section~\ref{sec2}. The experiment details are reported in Section~\ref{sec3}. Section~\ref{sec4} includes a discussion and the results for objective and subjective evaluation. Finally, the work is concluded in Section~\ref{sec5}.

\vspace{-0.15cm}
\section{CycleGAN for CLP Speech Intelligibility Improvement}
\label{sec2}

The CycleGAN is one of the adversarial networks that are widely used for VC or voice transformation. This section discusses the details of the CycleGAN system and its implementation for CLP speech intelligibility improvement in the following subsections. 


\subsection{Theory}

A CycleGAN consists of two generators $G$ and $F$ and two discriminators $D_X$ and $D_Y$, respectively. The generator $G$ is a function that maps the distribution $X$ into distribution $Y$, whereas the generator $F$ maps the distribution $Y$ into distribution $X$. On the other hand, the discriminator $D_X$ distinguishes $X$ from $\hat{X}=F(Y)$. In contrast, the discriminator $D_Y$ distinguishes $Y$ from $\hat{Y}=G(X)$. The CycleGAN model learns the mapping function from the training samples, which comprises of source $\{x_i\}^{N}_{i=1} \in X$ and target $\{y_i\}^{N}_{i=1} \in Y$ samples. The objective function of the CycleGAN model comprises of two losses: adversarial loss and cycle-consistency loss. An adversarial loss makes $X$ and $\hat{X}$ or $Y$ and $\hat{Y}$ as indistinguishable as possible. On the other hand, cycle-consistency loss guarantees that an input data retains its original characteristics after passing through the two generators. By combining both these losses (adversarial and cycle-consistency), a model can be learned from unpaired training data. The learned mappings can be further used to transform an input speech into the desired speech output. For adversarial loss, the objective function for mapping $G_{X\rightarrow Y}(x)$ and the corresponding discriminator $D_Y$ is given by, 
\begin{eqnarray}\label{eq1}
\begin{split}
\mathscr{L}_{GAN}(G,D_Y,X,Y)=E_{y \sim P (y)}[\log D_Y(y)] \\ 
+ E_{x \sim P (x)} [\log (1-D_Y(G(x)))]
\end{split}
\end{eqnarray}
where $P(x)$ and  $P(y)$ refer to the distribution of source and target data, respectively, and $E[\cdot]$ denotes the expectation operator. Using similar formulation as in Equation~(\ref{eq1}), the objective function for mapping $F_{Y\rightarrow X}(y)$ and corresponding discriminator $D_X$ is given by,
 
\begin{eqnarray}\label{eq2}
\begin{split}
	\mathscr{L}_{GAN}(F,D_X,X,Y)=E_{x \sim P (x)}[\log D_X(x)] \\ 
	+ E_{y \sim P (y)} [\log (1-D_X(F(y)))]
\end{split}
\end{eqnarray}
The generator attempts to generate data to minimize the two objective functions. At the same time, the discriminators $D_X$ and $D_Y$ try to maximize those two objective functions. Although the adversarial loss guarantees the distribution mapping, it does not guarantee that the learned function can map the input to the desired output. Furthermore, this may not serve the purpose of the current study, which is to improve the intelligibility of CLP speech while preserving the sequential information. Optimization of the adversarial loss does not guarantee linguistic consistency between input and output features. It is because adversarial loss only restricts the mapping function to follow target data distribution and does not necessarily retain the linguistic content of input speech. In order to address this issue, cycle-consistency loss is introduced in CycleGAN based VC, which finds the input and output pairs with same linguistic content. Therefore, the forward and backward cycle-consistency loss is given by,
 \begin{eqnarray}
 \label{eq3}
 \begin{split}
 \mathscr{L}_{cyc}(G,F)= E_{x \sim P (x)}[\left |  \right |F(G(x))-x\left |  \right |_1]\\
 + E_{y \sim P (y)} [\left |  \right |G(F(y))-y\left |  \right |_1]
 \end{split}
 \end{eqnarray}
where $\left |  \right |.\left |  \right |_1$ denotes $L1$ norm. Finally, the joint objective function to train CycleGAN is obtained by combining the adversarial loss with the cycle-consistent loss as given below,
 \begin{eqnarray}
 \label{eq4}
 \begin{split}
 \mathscr{L}(G,F,D_X,D_Y)=\mathscr{L}_{GAN}(G,D_Y,X,Y) \\ + \mathscr{L}_{GAN}(F,D_X,X,Y) + \lambda_{cyc}\mathscr{L}_{cyc}(G,F)
  \end{split}
  \end{eqnarray}
The $\lambda_{cyc}$ in Equation~(\ref{eq3}) controls the relative impact of adversarial loss and cycle-consistency loss. 

\subsection{System}

\begin{figure}[!t]
	{\centering
		{\includegraphics[width=0.48\textwidth]{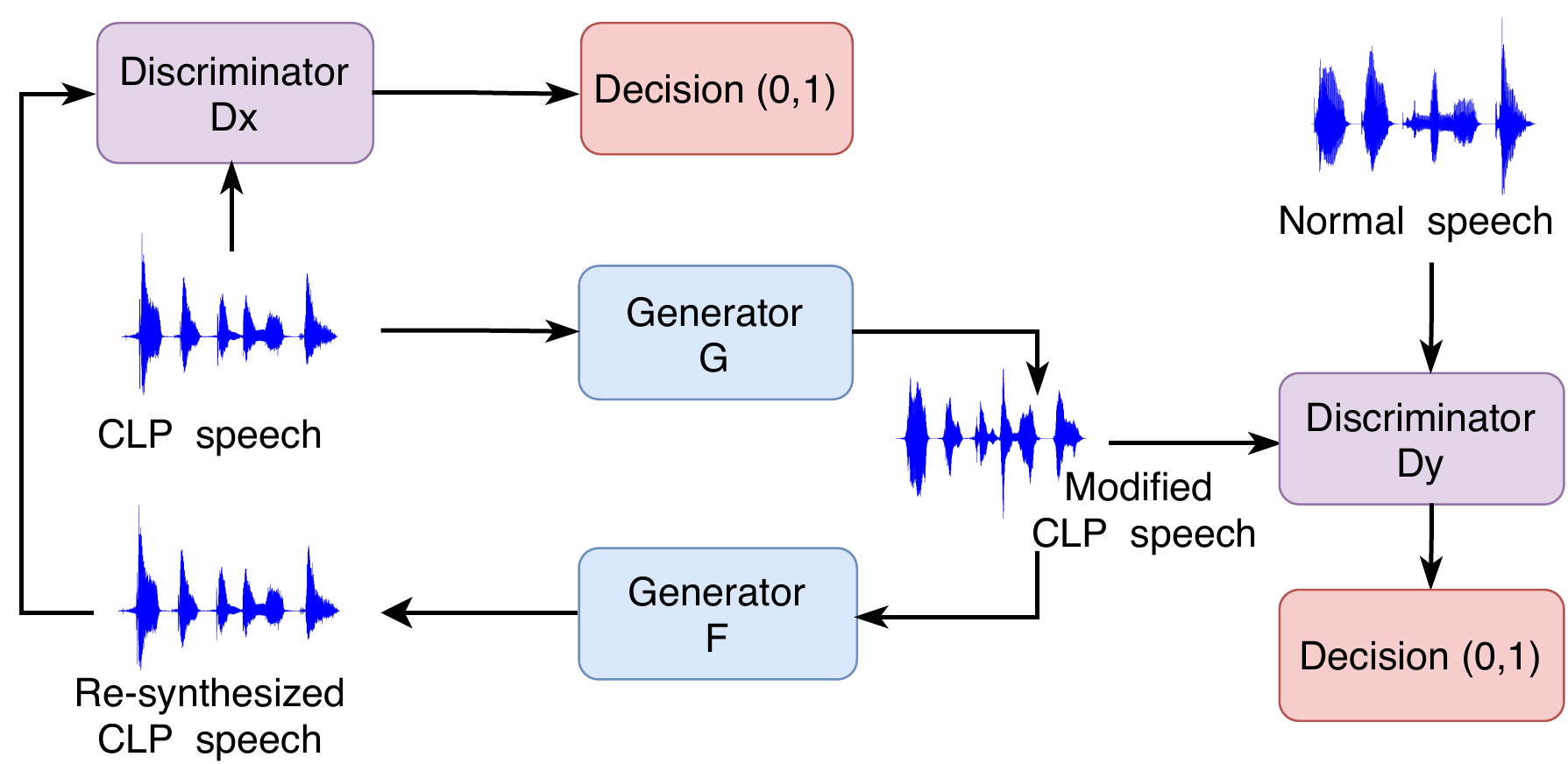} }
				\vspace{.2cm}
		\caption{Framework for the CLP speech enhancement using CycleGAN approach. }
		
		\label{framework}
	}
	\end{figure}

The focus of the current work is to enhance CLP speech $x \in X $ by mapping it to the normal signal $y \in Y $ without relying on parallel data. The transformation $G_{X \rightarrow Y}$ is performed using the CycleGAN method. Figure~\ref{framework} shows the framework for intelligibility enhancement for CLP speech using the CycleGAN system. The CLP speech serves as a source, whereas normal speech is considered as a target. Given a set of CLP and normal speech data, the CycleGAN learns a network to convert the CLP speech to normal speech, as discussed in the previous subsection. The discriminators and the generators work collectively during training. The generator serves as a mapping function from the distribution of the source to that of the target. On the other hand, the discriminator is trained to make the posterior probability 1 for normal speech and 0 for modified CLP speech. In contrast, the generator is trained to deceive the discriminator.


\vspace{-0.15cm}
\section{Experimental Setup}\label{sec3}
This section presents the details of the database and those of the experimental setup employed for this study.
\vspace{-0.15cm}
\subsection{Database}
The database used in this work consists of $62$ speakers, consisting of $31$ speakers ($17$ male and $14$ female) with CLP and $31$ non-CLP control speakers ($12$ male and $19$ female) in the Kannada language. The age of CLP and non-CLP participants are $9$ $\pm$ $2$ years (mean $\pm$ SD) and $10$ $\pm$ $2$ years (mean $\pm$ SD), respectively. The database consists of speech samples with disorders like hypernasality, articulation errors, and nasal air emission. It is noted that the manifestation of speech disorders are labeled by $3$ expert SLPs who have experience of not less than five years in the clinical field. The SLPs transcribed the speech samples and provide deviation scores on a scale of $0$ to $3$, where $0=$ close to normal, $1=$ mild deviation, $2=$ moderate deviation, and $3=$ severe deviation.

\begin{figure*}[t!]
	{\centering
		{\includegraphics[width=1\textwidth]{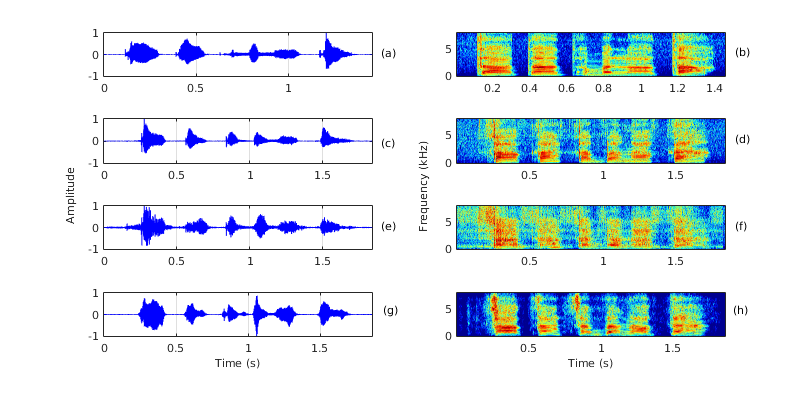} }
				\vspace{-0.9cm}
		\caption{ Waveforms and spectrograms of a sentence `taata tabala taa': (a) and (b) for normal speech; (c) and (d) for CLP speech; (e) and (f) for modified CLP speech using NMF; (g) and (h) for modified CLP speech using CycleGAN. }
	\vspace{-0.3cm}			
		\label{enhance}
	}
\end{figure*}

The database consists of sentences, vowel phonation, nonsense vowel-consonant-vowel (VCV), and consonant-vowel-consonant-vowel (CVCV) meaningful words. Only the sentences are used to train the CycleGAN model and perform speech intelligibility enhancement in the present work. Here, 5 CLP speakers (2 females and 3 males) are selected as sources, and 5 normal speakers (2 females and 3 males) are selected as targets. Each of these speakers has nineteen different spoken sentences recorded over two different sessions. This results in a total of 190 ($19 \times 2 \times 5 = 190$) sentences from the source speakers and the target speakers. The 190 sentences for CLP and normal speech are partitioned into 143 and 47 sentences as training and evaluation sets. There is no overlap among the sentences corresponding to the training and evaluation sets. 

\subsection{Implementation details}
The CycleGAN network architecture used in this work follows the one reported in the literature~\cite{kaneko2017sequence,kaneko2017generative}. The pipeline followed for CLP speech intelligibility enhancement includes the feature extraction, CycleGAN-based VC, and finally, re-synthesis of speech. The speech signals are downsampled to $16$~kHz for this study. The mel-cepstral coefficients extracted using the WORLD analysis system~\cite{morise2016world} are used as the features. The dimensionality of the feature vector is set to 24.

The CycleGAN model is collectively trained using the mel-cepstral coefficients extracted from each frame of the source and target speech data. Before processing, the source and target mel-cepstral coefficients are normalized to zero mean and unit variance. The training is stabilized using least-squares GAN~\cite{mao2017least}. It replaces the negative log-likelihood objective shown in Equation~(\ref{eq1}) by the least-squares loss. The $\lambda_{cyc}$ is set to 10. The randomness of each batch is increased using a fixed-length segment of 128 frames. We used Adam optimizer~\cite{kingma2014adam} to optimize the network with a batch size of 1. The generator and discriminator's initial learning rate is set as $0.0002$ and $0.0001$, respectively.

\section{Results and Discussion}
\label{sec4}


We now focus on the results of the studies conducted. The NMF based enhancement method is well explored in previous pathological speech studies~\cite{fu2017joint,aihara2012consonant,sudro2019modification}. Here, the speech signals are processed in a frame size of 20~ms and a shift of 10~ms. The speech sounds are characterized by 1024 fast-Fourier transform points. The magnitude spectrum is decomposed into a set of bases and nonnegative weights. The collection of bases is called a dictionary, and weights are referred to as activation. Before performing the conversion, the normal and CLP speech signals are time-aligned using the dynamic time warping method. It is followed by learning the two dictionaries simultaneously from the paired source and target training data. The source and target correspond to the normal and the CLP speech, respectively. The distorted speech is modified by using the target dictionary and shared activation matrix. 

Motivated by the wide use of the NMF method in different speech enhancement studies, it forms the baseline method for comparing the performance of CycleGAN based enhancement approach explored in this work. Before describing the details of objective and subjective evaluations, we would like to illustrate the differences between normal and CLP speech and the relative impact of NMF and CycleGAN processing on the modified CLP speech. For this purpose, the waveforms and spectrograms corresponding to normal, CLP, and the processed speech cases are shown in Figure~\ref{enhance}.


The CLP speech spectrogram in Figure~\ref{enhance}~(d) shows that the vowels are nasalized as compared to those of normal speech in Figure~\ref{enhance}~(b). The effect of nasalization is observed at around 1~kHz in between 0.5-0.7 seconds, between 1.0-1.4 seconds, and between 1.5-1.7 seconds. Additionally, the formants are not distinct in the vowels, and the stops are also observed to have been deviated in Figure~\ref{enhance}~(d) relative to those in Figure~\ref{enhance}~(b). The NMF based enhancement in Figure~\ref{enhance}~(f) shows that the nasalization is suppressed with distinctive formants. However, the deviant stop characteristics persist, and the speech is noisy. On observing the CycleGAN based enhancement in Figure~\ref{enhance}~(h), we find that the dominant low-frequency energy around 1~kHz is observed to be significantly suppressed, formants are distinct, and the stops are corrected. Thus, it reveals that the CycleGAN based enhanced CLP speech exhibits closer acoustic characteristics to that of the normal speech as compared to the NMF based one. For a thorough examination, some of the speech samples can be accessed using the link: \url{https://www.dropbox.com/sh/dpop7i7bhc3koig/AABQeUvl_v2telt70RV8H4Jra?dl=0}. Next, we report the objective and subjective evaluations in the following subsections. 


\vspace{-0.17cm}
\subsection{Objective Evaluation}
For this purpose, the intelligibility improvements of the modified speech signals are evaluated by ASR systems~\cite{ObjEval2020}. Two ASR systems are considered to evaluate the performance. The first one is based on publicly available open-source ASR for Indian English using Google API~\cite{google2020}. As the current study database is collected in the Kannada language, we also consider a Kannada ASR system trained using the KALDI speech recognition toolkit~\cite{povey2011kaldi} for evaluating the performance. The ASR system performance for various speech inputs is measured using the word error rate (WER) metric.


\begin{table}[t!]
	\centering
	\caption{Performance of different speech inputs using Google English ASR API and Kannada ASR system developed by the authors. The CLP$_{{\text{nmf}}}$ and CLP$_{{\text{cyclegan}}}$ denote modified CLP speech using NMF and CycleGAN, respectively.}
			\vspace{2mm}
	\label{asrg}
	\resizebox{8.5cm}{!}{
		\begin{tabular}{|c||c|c|c|c|}
			\hline
\multicolumn{1}{|c||}{\multirow{2}{*}{\bf ASR system}} & \multicolumn{4}{c|}{ \bf WER (\%)}                                                                                                       \\ \cline{2-5} 
\multicolumn{1}{|c||}{}                            & \multicolumn{1}{c|}{ \bf Normal} & \multicolumn{1}{c|}{\bf CLP} & \multicolumn{1}{c|}{\bf CLP$_{{\text{nmf}}}$} & \multicolumn{1}{c|}{\bf CLP$_{{\text {cyclegan}}}$} \\ \hline
			 \hline

Google English & 52.48        & 91.2       & 88.31       & 76.47     \\ \hline
		Kannada  &		24.03      &79.57       & 61.51       &{\bf 47.18}\\ \hline
		\end{tabular}
	}
	\vspace{-2mm}
\end{table}

Table~\ref{asrg} shows the ASR performance comparison of various speech inputs. It is observed that the performance of different speech inputs are better with the Kannada ASR system as there is a language match in contrast to the Google ASR system trained on Indian English. We note that the purpose of this study is not to compare the two ASR systems, but only to use Google ASR as another reference system to show the performance trend using a different recognizer, whose model is trained on a large dataset. As Google English API is readily available for use in the public domain; hence, for a sanity check, we have used it to show the performance accuracy. Any other API trained on a large dataset in any language could also be used in place of Google English API. 

Both the ASR systems are noted to exhibit severely degraded recognition performances for the CLP speech. However, both the explored enhancement approaches are noted to yield improved ASR performance compared to that of the original CLP speech, which is more prominent for the CycleGAN system. This signifies the motivation behind using CycleGAN for improving the intelligibility of CLP speech in the current study. 



\subsection{Subjective Evaluation}\label{seval}

In this subsection, we report the subjective evaluation based studies. This will lead us to have perceptual insights for the objective evaluation study presented in the previous subsection. A total of 20 listeners are chosen for this study. Each of them is provided ten sets of utterances that correspond to normal speech, CLP speech, and modified CLP speech using the NMF and CycleGAN approach, respectively. Each set is presented to the listeners without showing the speech file labels. 

\begin{figure}[t!]
	{\centering
\includegraphics[width=0.45\textwidth]{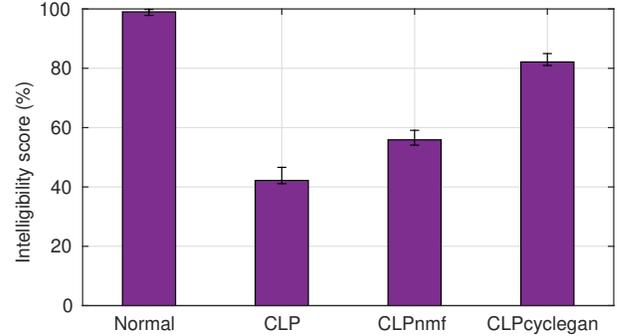}
		\caption{Bar plot showing intelligibility score (in \%) for different speech inputs evaluated by the naive listeners. The CLP$_{{\text{nmf}}}$ and CLP$_{{\text{cyclegan}}}$ denote modified CLP speech using NMF and CycleGAN, respectively.}
		\label{wer}
	}
\end{figure}

\begin{figure}[t!]
	{\centering
 				\includegraphics[width=0.45\textwidth]{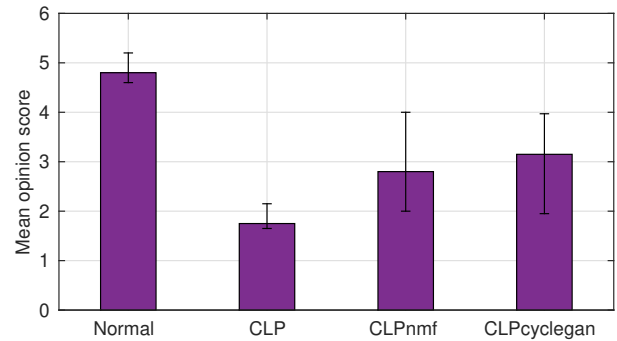}
 				
		\caption{Bar plot showing MOS for different speech inputs evaluated by the naive listeners. The CLP$_{{\text{nmf}}}$ and CLP$_{{\text{cyclegan}}}$ denote modified CLP speech using NMF and CycleGAN, respectively.}
				\vspace{-0.25cm}
		\label{mos}
	}
\end{figure}

The first study asks the listeners to transcribe the words spoken in those utterances based on their perception. We then compute the intelligibility score for the correctly recognized words for each speech category. The intelligibility score is evaluated in the range of 1-100\% based on the underlying spoken message perception. Figure~\ref{wer} shows the comparison of intelligibility scores conducted in this study for different speech. We observe that normal speech can be recognized correctly by the listeners showing a high intelligibility score. In contrast, it decreases to less than 50\% for CLP speech. The two enhancement methods (NMF and CycleGAN) improve the intelligibility, which is reflected in their respective gains, as observed from Figure~\ref{wer}. Further, it is found that CycleGAN can improve the intelligibility score above 80\%, thus showing effectiveness over the NMF approach.

The second subjective evaluation study is based on the mean opinion score (MOS) provided by the listeners for each utterance. The listeners are asked to rate the perceptual quality of speech on a scale of 1 to 5 (1 = bad, 2 = fair, 3 = good, 4 = very good, 5 = excellent). The average scores across all the listeners are computed to obtain the MOS for different speech categories. Figure~\ref{mos} shows the trend of MOS for this study. It can be noted that the normal speech has a MOS close to 5, whereas that of CLP speech is below 2. The speech signals obtained with the two enhancement methods NMF and CycleGAN, show an improved MOS than the original speech, which is higher for the latter approach. We obtain a MOS of more than 3 for the CycleGAN approach that shows the improvement in perceptual quality apart from intelligibility from the original CLP speech.  However, it is also observed that the modified CLP speech quality achieves a relatively lower MOS compared to that of the normal speech.

\section{Conclusions}
\label{sec5}
In this work, we study CycleGAN for enhancing the intelligibility of CLP speech. Through objective and subjective evaluation, it has been demonstrated that significant improvement in the intelligibility of CLP speech can be achieved using the CycleGAN based enhancement approach. The CLP speech enhanced using CycleGAN is noted to outperform the traditional NMF approach. It is worth pointing out that the differences in the perceptual quality of enhanced CLP and normal speech may be related to the vocoder and mapping parameters used, which deserves future exploration.

\section{Acknowledgements}

The authors would like to thank Dr. M. Pushpavathi and Dr. Ajish Abraham, AIISH Mysore, for providing insights about CLP speech disorder. The authors would also like to acknowledge the research scholars of IIT Guwahati for their participation in the subjective test. This work is, in part, supported by a project entitled “NASOSPEECH: Development of Diagnostic System for Severity Assessment of the Disordered Speech” funded by the Department of Biotechnology (DBT), Government of India. The work of the second author is also supported by Programmatic Grant No. A1687b0033 from the Singapore Government’s Research, Innovation and Enterprise 2020 plan (Advanced Manufacturing and Engineering domain).

\balance
\bibliographystyle{IEEEbib}
\bibliography{vwl_enh}

\end{document}